\documentclass[11pt]{article}
\usepackage[linesnumbered,ruled,vlined]{algorithm2e}
\usepackage{amsfonts}
\usepackage{amsmath}
\usepackage{amssymb}
\usepackage{authblk}
\usepackage{bm}
\usepackage{booktabs}
\usepackage[margin=1in]{geometry}
\usepackage{graphicx}
\usepackage[separate-uncertainty=true]{siunitx}
\usepackage[colorlinks=true]{hyperref}

\graphicspath{{figs/}}

\title{A comprehensive study of non-adaptive and residual-based adaptive sampling for physics-informed neural networks}
\author[1,\dag]{Chenxi Wu}
\author[1,\dag]{Min Zhu}
\author[2]{Qinyang Tan}
\author[3]{Yadhu Kartha}
\author[1,*]{Lu Lu}
\affil[1]{Department of Chemical and Biomolecular Engineering, University of Pennsylvania, Philadelphia, PA 19104, USA}
\affil[2]{Department of Mathematics, University of Southern California, Los Angeles, CA 90089, USA}
\affil[3]{College of Computing, Georgia Institute of Technology, Atlanta, GA 30332, USA}
\affil[$\dag$]{These authors contributed equally to this work.}
\affil[*]{Corresponding author. Email: lulu1@seas.upenn.edu}
\date{}

\begin{document}

\maketitle

\begin{abstract}
Physics-informed neural networks (PINNs) have shown to be an effective tool for solving both forward and inverse problems of partial differential equations (PDEs). PINNs embed the PDEs into the loss of the neural network using automatic differentiation, and this PDE loss is evaluated at a set of scattered spatio-temporal points (called residual points). The location and distribution of these residual points are highly important to the performance of PINNs. However, in the existing studies on PINNs, only a few simple residual point sampling methods have mainly been used. Here, we present a comprehensive study of two categories of sampling for PINNs: non-adaptive uniform sampling and adaptive nonuniform sampling. We consider six uniform sampling methods, including (1) equispaced uniform grid, (2) uniformly random sampling, (3) Latin hypercube sampling, (4) Halton sequence, (5) Hammersley sequence, and (6) Sobol sequence. We also consider a resampling strategy for uniform sampling. To improve the sampling efficiency and the accuracy of PINNs, we propose two new residual-based adaptive sampling methods: residual-based adaptive distribution (RAD) and residual-based adaptive refinement with distribution (RAR-D), which dynamically improve the distribution of residual points based on the PDE residuals during training. Hence, we have considered a total of 10 different sampling methods, including six non-adaptive uniform sampling, uniform sampling with resampling, two proposed adaptive sampling, and an existing adaptive sampling. We extensively tested the performance of these sampling methods for four forward problems and two inverse problems in many setups. Our numerical results presented in this study are summarized from more than \num{6000} simulations of PINNs. We show that the proposed adaptive sampling methods of RAD and RAR-D significantly improve the accuracy of PINNs with fewer residual points for both forward and inverse problems. The results obtained in this study can also be used as a practical guideline in choosing sampling methods.
\end{abstract}


\section{Introduction}

Physics-informed neural networks (PINNs)~\cite{raissi2019physics} have emerged in recent years and quickly became a powerful tool for solving both forward and inverse problems of partial differential equations (PDEs) via deep neural networks (DNNs)~\cite{raissi2020Hidden,lu2021deepxde,karniadakis2021physics}. PINNs embed the PDEs into the loss of the neural network using automatic differentiation. Compared with traditional numerical PDE solvers, such as the finite difference method (FDM) and the finite element method (FEM), PINNs are mesh free and therefore highly flexible. Moreover, PINNs can easily incorporate both physics-based constraints and data measurements into the loss function. PINNs have been applied to tackle diverse problems in computational science and engineering, such as inverse problems in nano-optics, metamaterials~\cite{chen2020physics}, and fluid dynamics~\cite{raissi2020Hidden}, parameter estimation in systems biology~\cite{yazdani2020systems,daneker2022identifiability}, and problems of inverse design and topology optimization~\cite{lu2021physics}. In addition to standard PDEs, PINNs have also been extended to solve other types of PDEs, including integro-differential equations~\cite{lu2021deepxde}, fractional PDEs~\cite{pang2019fPINN}, and stochastic PDEs~\cite{zhang2019quantifying}.

Despite the past success, addressing a wide range of PDE problems with increasing levels of complexity can be theoretically and practically challenging, and thus many aspects of PINNs still require further improvements to achieve more accurate prediction, higher computational efficiency, and training robustness~\cite{karniadakis2021physics}. A series of extensions to the vanilla PINN have been proposed to boost the performance of PINNs from various aspects. For example, better loss functions have been discovered via meta-learning~\cite{psaros2022meta}, and gradient-enhanced PINNs (gPINNs) have been developed to embed the gradient information of the PDE residual into the loss~\cite{yu2022Gradient}. In PINNs, the total loss is a weighted summation of multiple loss terms corresponding to the PDE and initial/boundary conditions, and different methods have been developed to automatically tune these weights and balance the losses~\cite{wang2021understanding,wang2022and,xiang2022self}. Moreover, a different weight for each loss term could be set at every training point~\cite{mcclenny2020self,gu2021selectnet,lu2021physics,li2022revisiting}. For problems in a large domain, decomposition of the spatio-temporal domain accelerates the training of PINNs and improves their accuracy~\cite{meng2020PPINN,shukla2021Parallel,Ameya2020XPINN}. For time-dependent problems, it is usually helpful to first train PINNs within a short time domain and then gradually expand the time intervals of training until the entire time domain is covered~\cite{wight2020solving,krishnapriyan2021characterizing,mattey2022novel,haitsiukevich2022improved,wang2022respecting}. In addition to these general methods, other problem-specific techniques have also been developed, e.g., enforcing Dirichlet or periodic boundary conditions exactly by constructing special neural network architectures~\cite{lagari2020systematic,dong2021method,lu2021physics}.

PINNs are mainly optimized against the PDE loss, which guarantees that the trained network is consistent with the PDE to be solved. PDE loss is evaluated at a set of scattered residual points. Intuitively, the effect of residual points on PINNs is similar to the effect of mesh points on FEM, and thus the location and distribution of these residual points should be highly important to the performance of PINNs. However, in previous studies on PINNs, two simple residual point sampling methods (i.e., an equispaced uniform grid and uniformly random sampling) have mainly been used, and the importance of residual point sampling has largely been overlooked.

\subsection{Related work and our contributions}

Different residual point sampling methods can be classified into two categories: uniform sampling and nonuniform sampling. Uniform sampling can be obtained in multiple ways. For example, we could use the nodes of an equispaced uniform grid as the residual points or randomly sample the points according to a continuous uniform distribution in the computational domain. Although these two sampling methods are simple and widely used, alternative sampling methods may be applied. The Latin hypercube sampling (LHS)~\cite{mckay2000comparison,stein1987large} was used in Ref.~\cite{raissi2019physics}, and the Sobol sequence~\cite{sobol1967distribution} was first used for PINNs in Ref.~\cite{pang2019fPINN}. The Sobol sequence is one type of quasi-random low-discrepancy sequences among other sequences, such as the Halton sequence~\cite{halton1960efficiency}, and the Hammersley sequence~\cite{hammersley1964monte}. Low-discrepancy sequences usually perform better than uniformly distributed random numbers in many applications such as numerical integration; hence, a comprehensive comparison of these methods for PINNs is required. However, very few comparisons~\cite{guo2020analysis,das2022state} have been performed. In this study, we
\begin{itemize}
    \item extensively compared the performance of different uniform sampling methods, including (1) equispaced uniform grid, (2) uniformly random sampling, (3) LHS, (4) Sobol sequence, (5) Halton sequence, and (6) Hammersley sequence.
\end{itemize}

In supervised learning, the dataset is fixed during training, but in PINNs, we can select residual points at any location. Hence, instead of using the same residual points during training, in each optimization iteration, we could select a new set of residual points, as first emphasized in Ref.~\cite{lu2021deepxde}. While this strategy has been used in some works, it has not yet been systematically tested. Thus, in this study, we
\begin{itemize}
    \item tested the performance of such a resampling strategy and investigated the effect of the number of residual points and the resampling period for the first time.
\end{itemize}

Uniform sampling works well for some simple PDEs, but it may not be efficient for those that are more complicated. To improve the accuracy, we could manually select the residual points in a nonuniform way, as was done in Ref.~\cite{mao2020physics} for high-speed flows, but this approach is highly problem-dependent and usually tedious and time-consuming. In this study, we focus on automatic and adaptive nonuniform sampling. Motivated by the adaptive mesh refinement in FEM, Lu et al.~\cite{lu2021deepxde} proposed the first adaptive nonuniform sampling for PINNs in 2019, the residual-based adaptive refinement (RAR) method, which adds new residual points in the locations with large PDE residuals. In 2021, another sampling strategy~\cite{nabian2021efficient} was developed, where all the residual points were resampled according to a probability density function (PDF) proportional to the PDE residual. In this study, motivated by these two ideas, we proposed two new sampling strategies:
\begin{itemize}
    \item residual-based adaptive distribution (RAD), where the PDF for sampling is a nonlinear function of the PDE residual;
    \item residual-based adaptive refinement with distribution (RAR-D), which is a hybrid method of RAR and RAD, i.e., the new residual points are added according to a PDF.
\end{itemize}
During the preparation of this paper, a few new studies appeared~\cite{zapf2022investigating,daw2022rethinking,gao2021active,tang2021deep,peng2022rang,zeng2022adaptive,hanna2022residual} that also proposed modified versions of RAR or PDF-based resampling. Most of these methods are special cases of the proposed RAD and RAR-D, and our methods can achieve better performance. We include a detailed comparison of these strategies in Section~\ref{sec:comparison}, after introducing several notations and our new proposed methods.

In this study, we have considered a total of 10 different sampling methods, including seven non-adaptive sampling methods (six different uniform samplings and one uniform sampling with resampling) and three adaptive sampling approaches (RAR, RAD, and RAR-D).
\begin{itemize}
    \item We compared the performance of these sampling methods for four forward problems of PDEs and investigated the effect of the number of residual points.
    \item We also compared their performance for two inverse problems that have not yet been considered in the literature.
    \item We performed more than 6000 simulations of PINNs to obtain all the results shown in this study.
\end{itemize}

\subsection{Organization}

This paper is organized as follows. In Section~\ref{sec:method}, after providing a brief overview of PINNs and different non-adaptive sampling strategies, two new adaptive nonuniform sampling strategies (RAD and RAR-D) are proposed. In Section~\ref{sec:results}, we compare the performance of 10 different methods for six different PDE problems, including four forward problems and two inverse problems. Section~\ref{sec:conclusion} summarizes the findings and concludes the paper.

\section{Methods}
\label{sec:method}

This section briefly reviews physics-informed neural networks (PINNs) in solving forward and inverse partial differential equations (PDEs). Then different types of uniformly sampling are introduced. Next, two nonuniform residual-based adaptive sampling methods are proposed to enhance the accuracy and training efficiency of PINNs. Finally, a comparison of related methods is presented.

\subsection{PINNs in solving forward and inverse PDEs}

We consider the PDE parameterized by $\bm{\lambda}$ defined on a domain $\Omega \subset \mathbb{R}^d$,
\begin{equation*}
    f(\mathbf{x}; u(\mathbf{x})) = f \left( \mathbf{x}; \frac{\partial u}{\partial x_{1}},...,\frac{\partial u}{\partial x_{d}};\frac{\partial^2 u}{\partial x_{1}\partial x_{1}},...,\frac{\partial^2 u}{\partial x_{1}\partial x_{d}};...;\bm{\lambda} \right) = 0, \quad \mathbf{x} = (x_{1},...,x_{d}) \in \Omega,
\end{equation*}
with boundary conditions on $\partial \Omega$
\begin{equation*}
    \mathcal{B}(u,\mathbf{x})=0,
\end{equation*}
and $u(\mathbf{x})$ denotes the solution at $\mathbf{x}$. In PINNs, the initial condition is treated as the Dirichlet boundary condition.

A forward problem is aimed to obtain the solution $u$ across the entire domain, where the model parameters $\bm{\lambda}$ are known. In practice, the model parameters $\bm{\lambda}$ might be unknown, but some observations from the solution $u$ are available, which lead to an inverse problem. An inverse problem is aimed to discover parameters $\bm{\lambda}$ that best describe the observed data from the solution.

PINNs are capable of addressing both forward and inverse problems. To solve a forward problem, the solution $u$ is represented with a neural network $\hat{u}(\mathbf{x}; \bm{\theta})$. The network parameters $\bm{\theta}$ are trained to approximate the solution $u$, such that the loss function is minimized~\cite{raissi2019physics,lu2021deepxde}:
\begin{equation*}
\mathcal{L}(\bm{\theta};\mathcal{T}) = w_f\mathcal{L}_f(\bm{\theta};\mathcal{T}_{f})+w_b\mathcal{L}_b(\bm{\theta};\mathcal{T}_{b}),
\end{equation*}
where
\begin{equation}\label{eq:loss_f}
\mathcal{L}_f(\bm{\theta};\mathcal{T}_{f}) = \frac{1}{|\mathcal{T}_f|}\sum_{\mathbf{x} \in \mathcal{T}_{f}} \left|f(\mathbf{x}; \frac{\partial \hat{u}} {\partial x_{1}},...,\frac{\partial \hat{u}}{\partial x_{d}};\frac{\partial^2 \hat{u}}{\partial x_{1}\partial x_{1}},...,\frac{\partial^2 \hat{u}}{\partial x_{1}\partial x_{d}};...;\bm{\lambda}) \right|^2,
\end{equation}
\begin{equation*}
\mathcal{L}_b(\bm{\theta};\mathcal{T}_{b}) = \frac{1}{|\mathcal{T}_b|} \sum_{\mathbf{x} \in \mathcal{T}_{b}} \left|\mathcal{B}(\hat{u},\mathbf{x})\right|^2,
\end{equation*}
and $w_f$ and $w_b$ are the weights.
Two sets of points are samples both inside the domain ($\mathcal{T}_{f}$) and on the boundaries ($\mathcal{T}_{b}$). Here, $\mathcal{T}_{f}$ and $\mathcal{T}_{b}$ are referred to as the sets of ``residual points'', and $\mathcal{T} = \mathcal{T}_{f} \cup \mathcal{T}_{b}$.

To solve the inverse problem, an additional loss term corresponding to the misfit of the observed data at the locations $\mathcal{T}_i$, defined as
\begin{equation*}
\mathcal{L}_i(\bm{\theta},\bm{\lambda};\mathcal{T}_{i}) = \frac{1}{|\mathcal{T}_i|}\sum_{\mathbf{x} \in \mathcal{T}_{i}}|\hat{u}(\mathbf{x})-u(\mathbf{x})|^2,
\end{equation*}
is added to the loss function. The loss function is then defined as
\begin{equation*}
\mathcal{L}(\bm{\theta},\bm{\lambda};\mathcal{T}) = w_f\mathcal{L}_f(\bm{\theta},\bm{\lambda};\mathcal{T}_{f})+w_b\mathcal{L}_b(\bm{\theta},\bm{\lambda};\mathcal{T}_{b})+w_i\mathcal{L}_i(\bm{\theta},\bm{\lambda};\mathcal{T}_{i}),
\end{equation*}
with an additional weight $w_i$. Then the network parameters $\bm{\theta}$ are trained simultaneously with $\bm{\lambda}$.

For certain PDE problems, it is possible to enforce boundary conditions directly by constructing a special network architecture~\cite{lagari2020systematic,dong2021method,lu2021physics,yu2022Gradient}, which eliminates the loss term of boundary conditions. In this study, the boundary conditions are enforced exactly and automatically. Hence, for a forward problem, the loss function is
\begin{equation*}
\mathcal{L}(\bm{\theta},\bm{\lambda};\mathcal{T}) = \mathcal{L}_f(\bm{\theta},\bm{\lambda};\mathcal{T}_{f}).
\end{equation*}
For an inverse problem, the loss function is
\begin{equation*}
\mathcal{L}(\bm{\theta},\bm{\lambda};\mathcal{T}) = w_f\mathcal{L}_f(\bm{\theta},\bm{\lambda};\mathcal{T}_{f})+w_i\mathcal{L}_i(\bm{\theta},\bm{\lambda};\mathcal{T}_{i}),
\end{equation*}
where we choose $w_f = w_i = 1$ for the diffusion-reaction equation in Section~\ref{sec:diffusion-reaction}, and $w_f = 1, w_i = 1000$ for the Korteweg-de Vries equation in Section~\ref{sec:kdv}.

\subsection{Uniformly-distributed non-adaptive sampling}

The training of PINNs requires a set of residual points ($\mathcal{T}_f$). The sampling strategy of $\mathcal{T}_f$ plays a vital role in promoting the accuracy and computational efficiency of PINNs. Here, we discuss several sampling approaches.

\subsubsection{Fixed residual points}
\label{sec:uniform_fixed}

In most studies of PINNs, we specify the residual points at the beginning of training and never change them during the training process. Two simple sampling methods (equispaced uniform grids and uniformly random sampling) have been commonly used. Other sampling methods, such as the Latin hypercube sampling (LHS)~\cite{mckay2000comparison,stein1987large} and the Sobol sequence~\cite{sobol1967distribution}, have also been used in some studies~\cite{raissi2019physics,pang2019fPINN,guo2020analysis}. The Sobol sequence is one type of quasi-random low-discrepancy sequences. Low-discrepancy sequences are commonly used as a replacement for uniformly distributed random numbers and usually perform better in many applications such as numerical integration. This study also considers other low-discrepancy sequences, including the Halton sequence~\cite{halton1960efficiency} and the Hammersley sequence~\cite{hammersley1964monte}.

\begin{figure}[htbp]
    \centering
    \includegraphics[width=\textwidth]{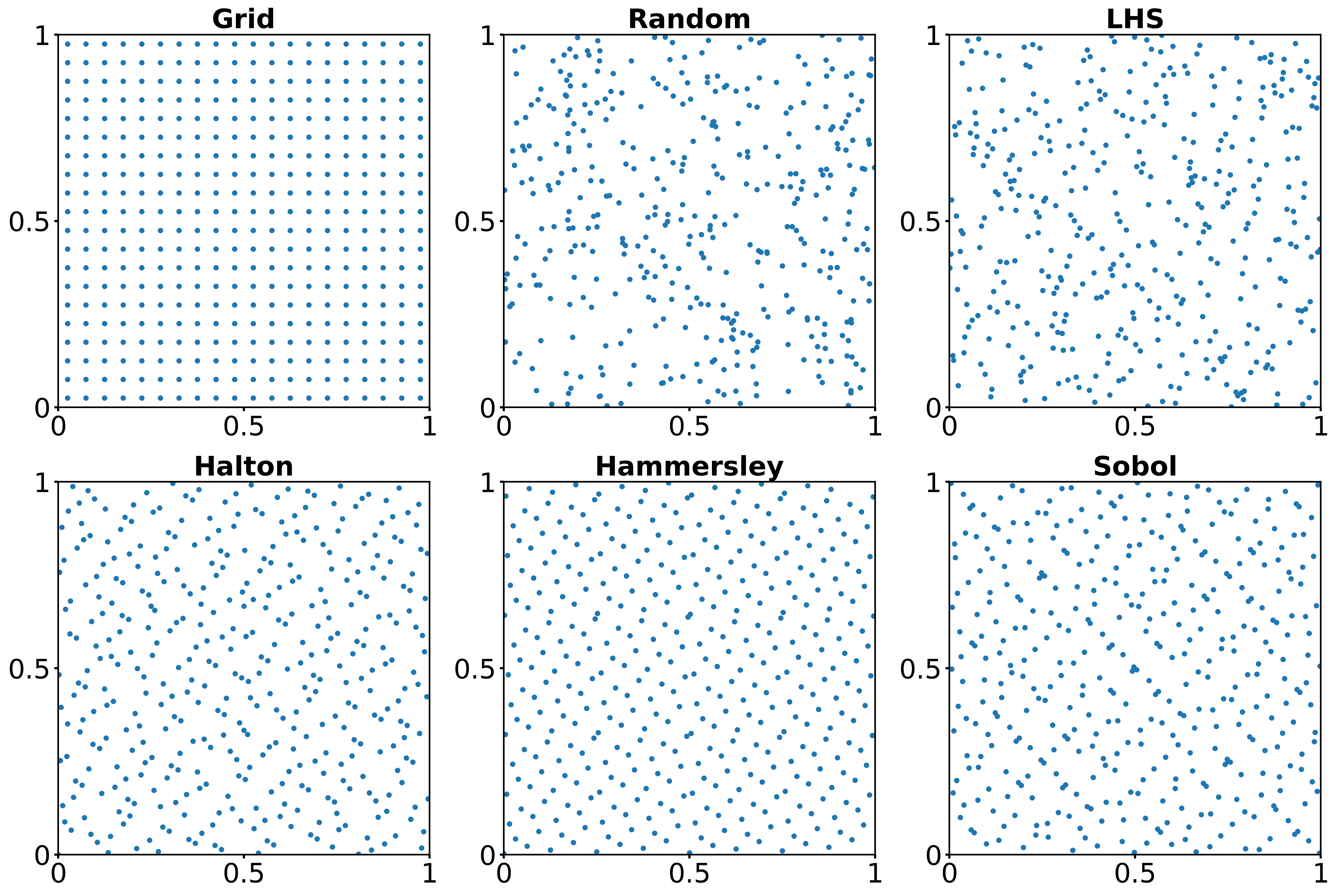}
    \caption{\textbf{Examples of 400 points generated in $[0,1]^2$ using different uniform sampling methods in Section~\ref{sec:uniform_fixed}.}}
    \label{fig:uniformpoints}
\end{figure}

We list the six uniform sampling methods as follows, and the examples of 400 points generated in $[0,1]^2$ using different methods are shown in Fig.~\ref{fig:uniformpoints}.
\begin{enumerate}
\item \textbf{Equispaced uniform grid (Grid)}: The residual points are chosen as the nodes of an equispaced uniform grid of the computational domain.
\item \textbf{Uniformly random sampling (Random)}: The residual points are randomly sampled according to a continuous uniform distribution over the domain. In practice, this is usually done using pseudo-random number generators such as the PCG-64 algorithm~\cite{o2014pcg}.
\item \textbf{Latin hypercube sampling (LHS)}~\cite{mckay2000comparison,stein1987large}: The LHS is a stratified Monte Carlo sampling method that generates random samples that occur within intervals on the basis of equal probability and with normal distribution for each range.
\item Quasi-random low-discrepancy sequences:
\begin{enumerate}
\item \textbf{Halton sequence (Halton)}~\cite{halton1960efficiency}: The Halton samples are generated according to the reversing or flipping the base conversion of numbers using primes.
\item \textbf{Hammersley sequence (Hammersley)}~\cite{hammersley1964monte}: The Hammersley sequence is the same as the Halton sequence, except in the first dimension where points are located equidistant from each other.
\item \textbf{Sobol sequence (Sobol)}~\cite{sobol1967distribution}: The Sobol sequence is a base-2 digital sequence that fills in a highly uniform manner.
\end{enumerate}
\end{enumerate}

\subsubsection{Uniform points with resampling}
\label{sec:random-r}

In PINNs, a point at any location can be used to evaluate the PDE loss. Instead of using the fixed residual points during training, we could also select a new set of residual points in every certain optimization iteration~\cite{lu2021deepxde}. The specific method to sample the points each time can be chosen from those methods discussed in Section~\ref{sec:uniform_fixed}. We can even use different sampling methods at different times, so many possible implementations make it impossible to be completely covered in this study.

In this study, we only consider Random sampling with resampling (\textbf{Random-R}). The Random-R method is the same as the Random method, except that the residual points are resampled for every $N$ iteration. The \textit{resampling period} $N$ is also an important hyperparameter for accuracy, as we demonstrate in our empirical experiments in Section~\ref{sec:results}.

\subsection{Nonuniform adaptive sampling}
\label{sec:adaptive_sampling}

Although the uniform sampling strategies were predominantly employed, recent studies on the nonuniform adaptive sampling strategies~\cite{lu2021deepxde,nabian2021efficient} have demonstrated promising improvement in the distribution of residual points during the training processes and achieved better accuracy.

\subsubsection{Residual-based adaptive refinement with greed (RAR-G)}

The first adaptive sampling method for PINNs is the residual-based adaptive refinement method (RAR) proposed in Ref.~\cite{lu2021deepxde}. RAR aims to improve the distribution of residual points during the training process by sampling more points in the locations where the PDE residual is large. Specifically, after every certain iteration, RAR adds new points in the locations with large PDE residuals (Algorithm~\ref{alg:rar-g}). RAR only focuses on the points with large residual, and thus it is a greedy algorithm. To better distinguish from the other sampling methods, the RAR method is referred to as RAR-G in this study.

\begin{algorithm}[htbp]
\caption{\textbf{RAR-G~\cite{lu2021deepxde}.}}
\label{alg:rar-g}
Sample the initial residual points $\mathcal{T}$ using one of the methods in Section~\ref{sec:uniform_fixed}\;
Train the PINN for a certain number of iterations\;
\Repeat{the total number of iterations or the total number of residual points reaches the limit}{
Sample a set of dense points $\mathcal{S}_0$ using one of the methods in Section~\ref{sec:uniform_fixed}\;
Compute the PDE residuals for the points in $\mathcal{S}_0$\;
$\mathcal{S} \gets$ $m$ points with the largest residuals in $\mathcal{S}_0$\;
$\mathcal{T} \gets \mathcal{T} \cup \mathcal{S}$\;
Train the PINN for a certain number of iterations\;
}
\end{algorithm}

\subsubsection{Residual-based adaptive distribution (RAD)}

RAR-G significantly improves the performance of PINNs when solving certain PDEs of solutions with steep gradients~\cite{lu2021deepxde,yu2022Gradient}. Nevertheless, RAR-G focuses mainly on the location where the PDE residual is largest and disregards the locations of smaller residuals. Another sampling strategy was developed later in Ref.~\cite{nabian2021efficient}, where all the residual points are resampled according to a probability density function (PDF) $p(\mathbf{x})$ proportional to the PDE residual. Specifically, for any point $\mathbf{x}$, we first compute the PDE residual $\varepsilon(\mathbf{x}) = \left|f(\mathbf{x}; \hat{u}(\mathbf{x}))\right|$, and then compute a probability as
\begin{equation*}
    p(\mathbf{x}) \propto \varepsilon(\mathbf{x}), \qquad \text{i.e.,} \qquad p(\mathbf{x}) = \frac{\varepsilon(\mathbf{x})}{A},
\end{equation*}
where $A = \int_\Omega \varepsilon(\mathbf{x}) dx$ is a normalizing constant. Then all the residual points are sampled according to $p(\mathbf{x})$.

This approach works for certain PDEs, but as we show in our numerical examples, it does not work well in some cases. Following this idea, we propose an improved version called the residual-based adaptive distribution (RAD) method (Algorithm~\ref{alg:rad}), where we use a new PDF defined as
\begin{equation}\label{eq:rad}
    p(\mathbf{x}) \propto \frac{\varepsilon^k(\mathbf{x})}{\mathbb{E}[ \varepsilon^k(\mathbf{x})]} + c,
\end{equation}
where $k \ge 0$ and $c \ge 0$ are two hyperparameters. $\mathbb{E}[ \varepsilon^k(\mathbf{x})]$ can be approximated by a numerical integration such as Monte Carlo integration. We note that the Random-R method in Section~\ref{sec:random-r} is a special case of RAD by choosing $k=0$ or $c \to \infty$.

\begin{algorithm}[htbp]
\caption{\textbf{RAD.}}
\label{alg:rad}
Sample the initial residual points $\mathcal{T}$ using one of the methods in Section~\ref{sec:uniform_fixed}\;
Train the PINN for a certain number of iterations\;
\Repeat{the total number of iterations reaches the limit}{
$\mathcal{T} \gets$ A new set of points randomly sampled according to the PDF of Eq.~\eqref{eq:rad}\;
Train the PINN for a certain number of iterations\;
}
\end{algorithm}

In RAD (Algorithm~\ref{alg:rad} line 4), we need to sample a set of points according to $p(\mathbf{x})$, which can be done in a few ways. When $\mathbf{x}$ is low-dimensional, we can sample the points approximately in the following brute-force way:
\begin{enumerate}
    \item Sample a set of dense points $\mathcal{S}_0$ using one of the methods in Section~\ref{sec:uniform_fixed};
    \item Compute $p(\mathbf{x})$ for the points in $\mathcal{S}_0$;
    \item Define a probability mass function $\tilde{p}(\mathbf{x}) = \frac{p(\mathbf{x})}{A}$ with the normalizing constant $A = \sum_{\mathbf{x} \in \mathcal{S}_0} p(\mathbf{x})$;
    \item Sample a subset of points from $\mathcal{S}_0$ according to $\tilde{p}(\mathbf{x})$.
\end{enumerate}
This method is simple, easy to implement, and sufficient for many PDE problems. For more complicated cases, we can use other methods such as inverse transform sampling, Markov chain Monte Carlo (MCMC) methods, and generative adversarial networks (GANs)~\cite{goodfellow2014generative}.

\begin{figure}[htbp]
    \centering
    \includegraphics[width=\textwidth]{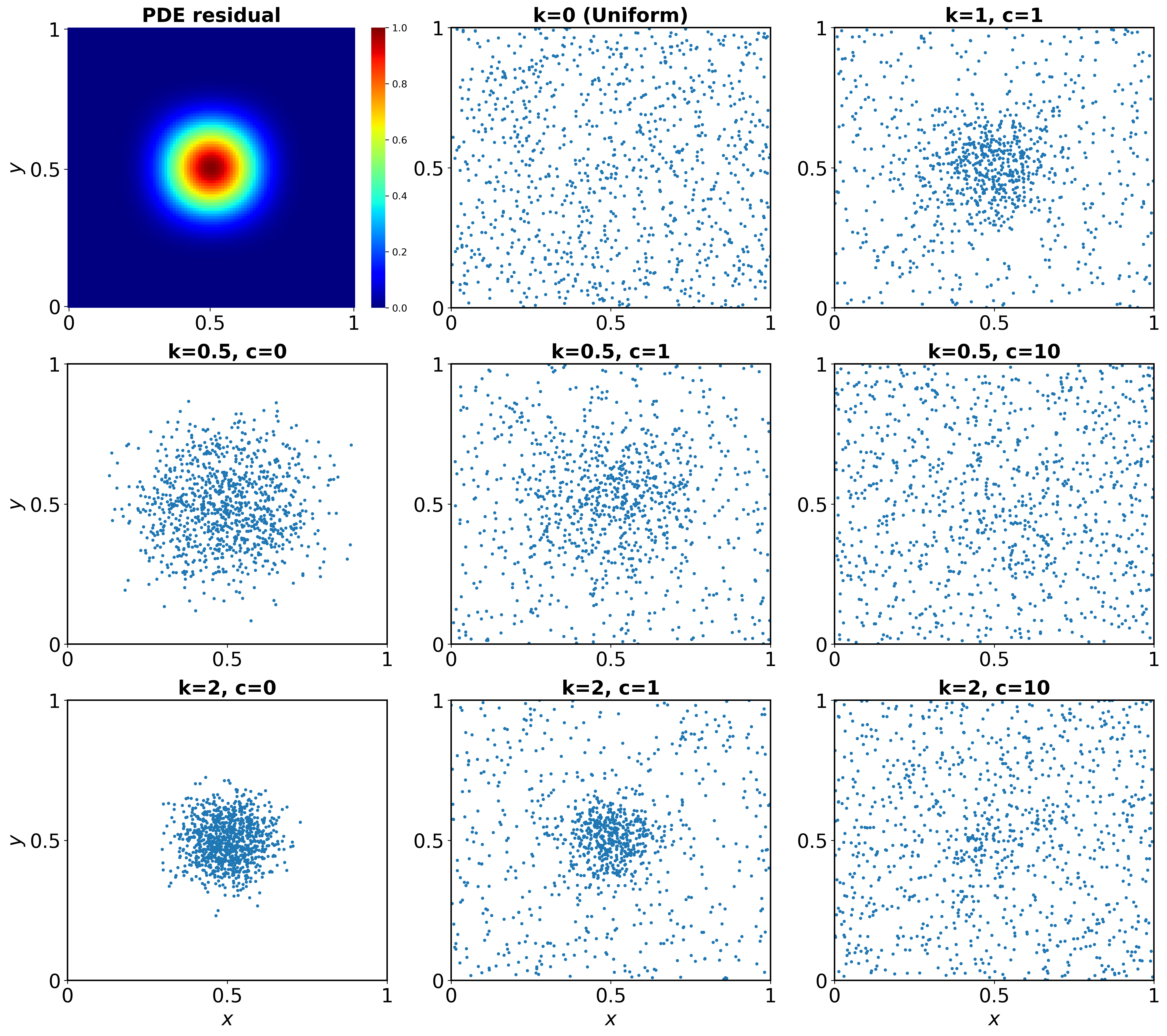}
    \caption{\textbf{Examples of 1000 residual points sampled by RAD with different values of $k$ and $c$ for the PDE residual $\varepsilon(x,y)$ in Eq.~\eqref{eq:rad_ex}.}}
    \label{fig:RAD}
\end{figure}

The two hyperparameters $k$ and $c$ in Eq.~\eqref{eq:rad} control the profile of $p(\mathbf{x})$ and thus the distribution of sampled points. We illustrate the effect of $k$ and $c$ using a simple 2D example,
\begin{equation}\label{eq:rad_ex}
    \varepsilon(x,y) = 2^{4a} x^a(1-x)^ay^a(1-y)^a,
\end{equation}
with $a=10$ in Fig.~\ref{fig:RAD}. When $k=0$, it becomes a uniform distribution. As the value of $k$ increases, more residual points will large PDE residuals are sampled. As the value of $c$ increases, the residual points exhibit an inclination to be uniformly distributed. Compared with RAR, RAD provides more freedom to balance the points in the locations with large and small residuals by tuning $k$ and $c$. The optimal values of $k$ and $c$ are problem-dependent, and based on our numerical results, the combination of $k=1$ and $c=1$ is usually a good default choice.

\subsubsection{Residual-based adaptive refinement with distribution (RAR-D)}

We also propose a hybrid method of RAR-G and RAD, namely, residual-based adaptive refinement with distribution (RAR-D) (Algorithm~\ref{alg:rar-d}). Similar to RAR-G, RAR-D repeatedly adds new points to the training dataset; similar to RAD, the new points are sampled based on the PDF in Eq.~\eqref{eq:rad}. We note that when $k \to \infty$, only points with the largest PDE residual are added, which recovers RAR-G. The optimal values of $k$ and $c$ are problem dependent, and based on our numerical results, the combination of $k=2$ and $c=0$ is usually a good default choice.

\begin{algorithm}[htbp]
\caption{\textbf{RAR-D.}}
\label{alg:rar-d}
Sample the initial residual points $\mathcal{T}$ using one of the methods in Section~\ref{sec:uniform_fixed}\;
Train the PINN for a certain number of iterations\;
\Repeat{the total number of iterations or the total number of residual points reaches the limit}{
$\mathcal{S} \gets$ $m$ points randomly sampled according to the PDF of Eq.~\eqref{eq:rad}\;
$\mathcal{T} \gets \mathcal{T} \cup \mathcal{S}$\;
Train the PINN for a certain number of iterations\;
}
\end{algorithm}

\subsection{Comparison with related work}
\label{sec:comparison}

As discussed in Section~\ref{sec:adaptive_sampling}, our proposed RAD and RAR-D are improved versions of the methods in Refs.~\cite{lu2021deepxde,nabian2021efficient}. Here, we summarize the similarities between their methods and ours.
\begin{itemize}
    \item Lu et al.~\cite{lu2021deepxde} (in July 2019) proposed RAR (renamed to RAR-G here), which is a special case of RAR-D by choosing a large value of $k$.
    \item The method proposed by Nabian et al.~\cite{nabian2021efficient} (in April 2021) is a special case of RAD by choosing $k=1$ and $c=0$.
\end{itemize}

During the preparation of this paper, a few new papers appeared~\cite{zapf2022investigating,daw2022rethinking,gao2021active,tang2021deep,peng2022rang,zeng2022adaptive,hanna2022residual} that also proposed similar methods. Here, we summarize the similarities and differences between these studies.
\begin{itemize}
    \item The method proposed by Gao et al.~\cite{gao2021active} (in December 2021) is a special case of RAD by choosing $c=0$.
    \item Tang et al.~\cite{tang2021deep} (in December 2021) proposed two methods. One is a special case of RAD by choosing $k=2$ and $c=0$, and the other is a special case of RAR-D by choosing $k=2$ and $c=0$. 
    \item Zeng et al.~\cite{zeng2022adaptive} (in April 2022) proposed a subdomain version of RAR-G. The entire domain is divided into many subdomains, and then new points are added to the several subdomains with large average PDE residual.
    \item Similar to RAR-G, Peng et al.~\cite{peng2022rang} (in May 2022) proposed to add more points with large PDE residual, but they used the node generation technology proposed in Ref.~\cite{fornberg2015fast}. We note that this method only works for a two-dimensional space.
    \item Zapf et al.~\cite{zapf2022investigating} (in May 2022) proposed a modified version of RAR-G, where some points with small PDE residual are removed while adding points with large PDE residual. They show that compared with RAR, this reduces the computational cost, but the accuracy keeps similar.
    \item Hanna et al.~\cite{hanna2022residual} (in May 2022) proposed a similar method as RAR-D, but they chose $p(\mathbf{x}) \propto \max\{\log(\varepsilon(\mathbf{x})/\varepsilon_0), 0\}$, where $\varepsilon_0$ is a small tolerance.
    \item Similar to the work of Zapf et al., Daw et al.~\cite{daw2022rethinking} (in July 2022) also proposed to remove the points with small PDE residual, but instead of adding new points with large PDE residual, they added new uniformly random sampled points. 
\end{itemize}

Thus all these methods are special cases of our proposed RAD and RAR-D (or with minor modification). However, in our study, two tunable variables $k$ and $c$ are introduced. As we show in our results, the values of $k$ and $c$ could be crucial since they significantly influence the residual points distribution. By choosing proper values of $k$ and $c$, our methods would outperform the other methods.

We also note that the point-wise weighting~\cite{mcclenny2020self,gu2021selectnet,lu2021physics,li2022revisiting} can be viewed as a special case of adaptive sampling, described as follows. When the residual points are randomly sampled from a uniform distribution $\mathcal{U}(\Omega)$, and the number of residual points is large, the PDE loss in Eq.~\eqref{eq:loss_f} can be approximated by $\mathbb{E}_\mathcal{U} [\varepsilon^2(\mathbf{x})]$. If we consider a point-wise weighting function $w(\mathbf{x})$, then the loss becomes $\mathbb{E}_\mathcal{U} [w(\mathbf{x}) \varepsilon^2(\mathbf{x})]$, while for RAD the loss is $\mathbb{E}_p [\varepsilon^2(\mathbf{x})]$. If we choose $w(\mathbf{x})$ (divided by a normalizing constant) as the PDF $p(\mathbf{x})$, then the two losses are equal.

\section{Results}
\label{sec:results}

We apply PINNs with all the ten sampling methods in Section~\ref{sec:method} to solve six forward and inverse PDE problems. In all examples, the hyperbolic tangent ($\tanh$) is selected as the activation function. Table~\ref{tab:hyperparameters} summarizes the network width, depth, and optimizers used for each example. More details of the hyperparameters and training procedure can be found in each section of the specific problem.

\begin{table}[htbp]
\caption{\textbf{The hyperparameters used for each numerical experiment.} The learning rate of Adam optimizer is chosen as 0.001.}
\label{tab:hyperparameters}
\centering
\begin{tabular}{l|ccc}
  \toprule
  Problems   & Depth    & Width &   Optimizer \\
  \midrule
  Section~\ref{sec:diffusion} Diffusion equation &   4  & 32   &Adam \\
  Section~\ref{sec:burgers} Burgers' equation &   4  & 64   & Adam + L-BFGS \\
  Section~\ref{sec:allencahn} Allen-Cahn equation &   4  & 64   & Adam + L-BFGS \\
  Section~\ref{sec:wave} Wave equation &   6  & 100  & Adam + L-BFGS \\
  Section~\ref{sec:diffusion-reaction} Diffusion-reaction equation (inverse) &   4  & 20   & Adam \\
  Section~\ref{sec:kdv} Korteweg-de Vries equation (inverse) &   4  & 100  & Adam \\
  \bottomrule
\end{tabular}
\end{table}

For both forward and inverse problems, to evaluate the accuracy of the solution $\hat{u}$, the $L^2$ relative error is used:
\begin{equation*}
\frac{\|\hat{u}-u\|_{2}}{\|u\|_{2}}.
\end{equation*}
For inverse problems, to evaluate the accuracy of the predicted coefficients $\hat{\lambda}$, the relative error is also computed:
\begin{equation*}
\frac{|\hat{\lambda}-\lambda|}{|\lambda|}.
\end{equation*}
As the result of PINN has randomness due to the random sampling, network initialization, and optimization, thus, for each case, we run the same experiment at least 10 times and then compute the geometric mean and standard deviation of the errors. The code in this study is implemented by using the library DeepXDE~\cite{lu2021deepxde} and is publicly available from the GitHub repository \url{https://github.com/lu-group/pinn-sampling}.

\subsection{Summary}
\label{sec:summary}

Here, we first present a summary of the accuracy of all the methods for the forward and inverse problems listed in Tables~\ref{tab:error-forward} and Table~\ref{tab:error-inverse}, respectively. A relatively small number of residual points is chosen to show the difference among different methods. In the specific section of each problem (Sections~\ref{sec:diffusion}--\ref{sec:kdv}), we discuss all the detailed analyses, including the convergence of error during the training process, the convergence of error with respect to the number of residual points, and the effects of different hyperparameters (e.g., the period of resampling in Random-R, the values of $k$ and $c$ in RAD and RAR-D, and the number of new points added each time in RAR-D). We note that Random-R is a special case of RAD by choosing $k=0$ or $c \to \infty$, and RAR-G is a special case of RAR-D by choosing $k \to \infty$.

Our main findings from the results are as follows.
\begin{itemize}
    \item The proposed RAD method has always performed the best among the 10 sampling methods when solving all forward and inverse problems.
    \item For PDEs with complicated solutions, such as the Burgers' and multi-scale wave equation, the proposed RAD and RAR-D methods are predominately effective and yield errors magnitudes lower.
    \item For PDEs with smooth solutions, such as the diffusion equation and diffusion-reaction equation, some uniform sampling methods, such as the Hammersley and Random-R, also produce sufficiently low errors.
    \item Compared with other uniform sampling methods, Random-R usually demonstrates better performance.
    \item Among the six uniform sampling methods with fixed residual points, the low-discrepancy sequences (Halton, Hammersley, and Sobol) generally perform better than Random and LHS, and both are better than Grid.
\end{itemize}

\begin{table}[htbp]
    \centering
    \caption{\textbf{$L^2$ relative error of the PINN solution for the forward problems.} Bold font indicates the smallest three errors for each problem. Underlined text indicates the smallest error for each problem.}
    \label{tab:error-forward}
    \begin{tabular}{l|cccc}
    \toprule
    & Diffusion & Burgers' & Allen-Cahn & Wave \\
    \midrule
    No. of residual points & 30 & 2000 & 1000 & 2000 \\
    \midrule
    Grid & 0.66 $\pm$ 0.06\% & 13.7 $\pm$ 2.37\%  & 93.4 $\pm$ 6.98\% & 81.3 $\pm$ 13.7\%\\
    Random & 0.74 $\pm$ 0.17\% & 13.3 $\pm$ 8.35\% & 22.2 $\pm$ 16.9\% & 68.4 $\pm$ 20.1\%\\
    LHS & 0.48 $\pm$ 0.24\% & 13.5 $\pm$ 9.05\% & 26.6 $\pm$ 15.8\% & 75.9 $\pm$ 33.1\%\\
    Halton & 0.24 $\pm$ 0.17\% & 4.51 $\pm$ 3.93\% & 0.29 $\pm$ 0.14\% & 60.2 $\pm$ 10.0\%\\
    Hammersley & 0.17 $\pm$ 0.07\% & 3.02 $\pm$ 2.98\% & \textbf{0.14 $\pm$ 0.14}\% & 58.9 $\pm$ 8.52\%\\
    Sobol & 0.19 $\pm$ 0.07\% & 3.38 $\pm$ 3.21\% & 0.35 $\pm$ 0.24\% & 57.5 $\pm$ 14.7\% \\
    \midrule
    Random-R & \textbf{0.12 $\pm$ 0.06}\% & 1.69 $\pm$ 1.67\% & 0.55 $\pm$ 0.34\% & \textbf{0.72 $\pm$ 0.90}\% \\
    \midrule
    RAR-G~\cite{lu2021deepxde} & 0.20 $\pm$ 0.07\% & \textbf{0.12 $\pm$ 0.04}\%  & 0.53 $\pm$ 0.19\% & 0.81 $\pm$ 0.11\%\\
    RAD & \underline{\textbf{0.11 $\pm$ 0.07}}\% & \underline{\textbf{0.02 $\pm$ 0.00}}\% & \underline{\textbf{0.08 $\pm$ 0.06}}\% & \underline{\textbf{0.09 $\pm$ 0.04}}\%\\
    RAR-D & \textbf{0.14 $\pm$ 0.11}\% & \textbf{0.03 $\pm$ 0.01}\% & \textbf{0.09 $\pm$ 0.03}\% & \textbf{0.29 $\pm$ 0.04}\% \\
    \bottomrule
    \end{tabular}
\end{table}

\begin{table}[htbp]
    \centering
    \caption{\textbf{$L^2$ relative error of the PINN solution and relative error of the inferred parameters for the inverse problems.} Bold font indicates the smallest three errors for each problem. Underlined text indicates the smallest error for each problem.}
    \label{tab:error-inverse}
    \begin{tabular}{l|cc|ccc}
    \toprule
    & \multicolumn{2}{c}{Diffusion-reaction} & \multicolumn{3}{c}{Korteweg-de Vries} \\
    & $u(x)$ & $k(x)$ & $u(x,t)$ & $\lambda_1$ & $\lambda_2$ \\
    \midrule
    No. of residual points & \multicolumn{2}{c}{15} & \multicolumn{3}{c}{600} \\
    \midrule
    Grid & 0.36 $\pm$ 0.12\% & 8.58 $\pm$ 2.14\% & 24.4 $\pm$ 11.1\% & 53.7 $\pm$ 30.7\% & 42.0 $\pm$ 22.3\% \\
    Random & 0.35 $\pm$ 0.17\% & 5.77 $\pm$ 2.05\% & 8.86 $\pm$ 2.80\% & 16.4 $\pm$ 7.33\% & 16.8 $\pm$ 7.40\% \\
    LHS & 0.36 $\pm$ 0.14\% & 7.00 $\pm$ 2.62\% & 10.9 $\pm$ 2.60\% & 22.0 $\pm$ 6.68\% & 22.6 $\pm$ 6.36\% \\
    Halton & 0.23 $\pm$ 0.08\% & 6.16 $\pm$ 1.08\% & 8.76 $\pm$ 3.33\% & 16.7 $\pm$ 6.16\% & 17.2 $\pm$ 6.20\% \\
    Hammersley & 0.28 $\pm$ 0.08\% & 6.37 $\pm$ 0.91\% & 4.49 $\pm$ 3.56\% & 5.24 $\pm$ 7.08\% & 5.71 $\pm$ 7.32\% \\
    Sobol & \textbf{0.21 $\pm$ 0.06}\% & \textbf{3.09 $\pm$ 0.75}\% & 8.59 $\pm$ 3.67\% & 15.8 $\pm$ 6.15\% & 15.6 $\pm$ 5.79\% \\
    \midrule
    Random-R & \textbf{0.19 $\pm$ 0.09}\% & \textbf{3.43 $\pm$ 1.80}\% & \textbf{0.97 $\pm$ 0.15}\% & \textbf{0.41 $\pm$ 0.30}\% & \textbf{1.14 $\pm$ 0.31}\% \\
    \midrule
    RAR-G~\cite{lu2021deepxde} & 1.12 $\pm$ 0.11\% & 15.9 $\pm$ 1.53\% & 8.83 $\pm$ 1.98\% & 15.4 $\pm$ 9.29\% & 14.5 $\pm$ 9.25\% \\
    RAD & \underline{\textbf{0.17 $\pm$ 0.09}}\% & \underline{\textbf{2.76 $\pm$ 1.32}}\% & \underline{\textbf{0.77 $\pm$ 0.11}}\% & \underline{\textbf{0.31 $\pm$ 0.19}}\% & \underline{\textbf{0.86 $\pm$ 0.25}}\% \\
    RAR-D & 0.76 $\pm$ 0.24\% & 10.3 $\pm$ 3.28\% & \textbf{2.36 $\pm$ 0.98}\% & \textbf{3.49 $\pm$ 2.21}\% & \textbf{3.18 $\pm$ 2.02}\% \\
    \bottomrule
    \end{tabular}
\end{table}

\subsection{Diffusion equation}
\label{sec:diffusion}

We first consider the following one-dimensional diffusion equation:
\begin{gather*}
    \frac{\partial u}{\partial t} =  \frac{\partial^2 u}{\partial x^2} + e^{-t}\left(-\sin (\pi x) + \pi^2\sin(\pi x)\right), \quad x \in [-1,1], t \in [0,1], \\
    u(x, 0) = \sin (\pi x), \\
    u(-1, t) = u(1, t) = 0,
\end{gather*}
where $u$ is the concentration of the diffusing material. The exact solution is $u(x,t) = \sin (\pi x) e^{-t}$.

\begin{figure}[htbp]
    \centering
    \includegraphics[scale=0.28]{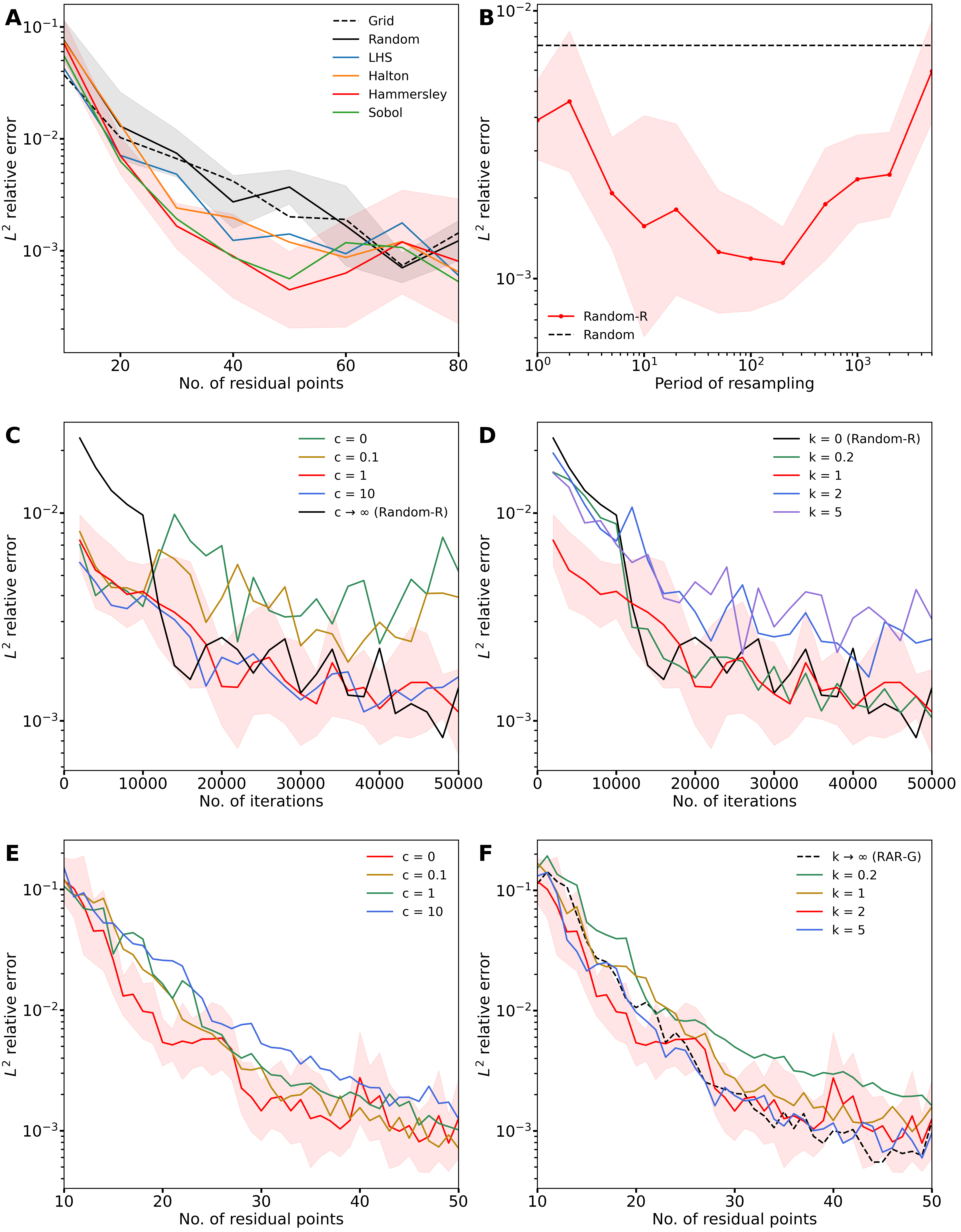}
    \caption{\textbf{$L^2$ relative errors of different sampling methods for the diffusion equation in Section~\ref{sec:diffusion}.} (\textbf{A}) Six uniform sampling with fixed residual points. (\textbf{B}) Random-R with different periods of resampling when using 30 residual points. (\textbf{C} and \textbf{D}) The training trajectory of RAD with different values of $k$ and $c$ when using 30 residual points. (C) $k = 1$. (D) $c = 1$. (\textbf{E} and \textbf{F}) RAR-D with different values of $k$ and $c$. Each time one new point is added. (E) $k$ = 2. (F) $c$ = 0. The curves and shaded regions represent the geometric mean and one standard deviation of 10 runs. For clarity, only some standard deviations are plotted.}
    \label{fig:diffusion}
\end{figure}

We first compare the performance of the six uniform sampling methods with fixed residual points (Fig.~\ref{fig:diffusion}A). The number of residual points is ranged from 10 to 80 with an increment of 10 points each time. For each number of residual points, the maximum iteration is set to be \num{15000} with Adam as the optimizer. When the number of points is large (e.g., more than 70), all these methods have similar performance. However, when the number of residual points is small such as 50, the Hammersley and Sobol sequences perform better than others, and the equispaced uniform grid and random sampling have the largest errors (about one order of magnitude larger than Hammersley and Sobol).

We then test the Random-R method using 30 residual points (Fig.~\ref{fig:diffusion}B). The accuracy of Random-R has a strong dependence on the period of resampling, and the optimal period of resampling in this problem is around 200. Compared with Random without resampling, the Random-R method always leads to lower $L^2$ relative errors regardless of the period of resampling. The error can be lower by one order of magnitude by choosing a proper resampling period. Among all the non-adaptive methods, Random-R performs the best.

Next, we test the performance of the nonuniform adaptive sampling methods. In Algorithms~\ref{alg:rad} and \ref{alg:rar-d}, the neural network is first trained using \num{10000} steps of Adam. In the RAD method, we use 30 residual points and resample every \num{1000} iterations. The errors of RAD with different values of $k$ and $c$ are shown in Figs.~\ref{fig:diffusion}C and D. We note that Random-R is a special case of RAD with either $c \to \infty$ or $k=0$. Here, RAD with large values of $c$ or small values of $k$ leads to better accuracy, i.e., the points are almost uniformly distributed. For the RAR-D method (Figs.~\ref{fig:diffusion}E and F), one residual point is added after every \num{1000} iterations starting from 10 points. When using $k=2$ and $c=0$ (the two red lines in Figs.~\ref{fig:diffusion}E red F), RAR-D performs the best.

When using 30 residual points, the errors of all the methods are listed in Table~\ref{tab:error-forward}. In this diffusion equation, all the methods achieve a good accuracy ($<1\%$). Compared with Random-R (0.12\%), RAD and RAR-D (0.11\%) are not significantly better. The reason could be that the solution of this diffusion equation is very smooth, so uniformly distributed points are good enough. In our following examples, we show that RAD and RAR-D work significantly better and achieve an error of orders of magnitude smaller than the non-adaptive methods.

\subsection{Burgers' equation}
\label{sec:burgers}

The Burgers' equation is considered defined as:
\begin{gather*}
    \frac{\partial u}{\partial t} + u \frac{\partial u}{\partial x} = \nu \frac{\partial^2 u}{\partial x^2}, \quad x \in [-1,1], t \in [0,1], \\
    u(x, 0) = -\sin (\pi x), \\
    u(-1, t) = u(1, t) = 0,
\end{gather*}
where $u$ is the flow velocity and $\nu$ is the viscosity of the fluid. In this study, $\nu$ is set at $0.01/\pi$. Different from the diffusion equation with a smooth solution, the solution of the Burgers' equation has a sharp front when $x=0$ and $t$ is close to 1.

\begin{figure}[htbp]
    \centering
    \includegraphics[scale=0.28]{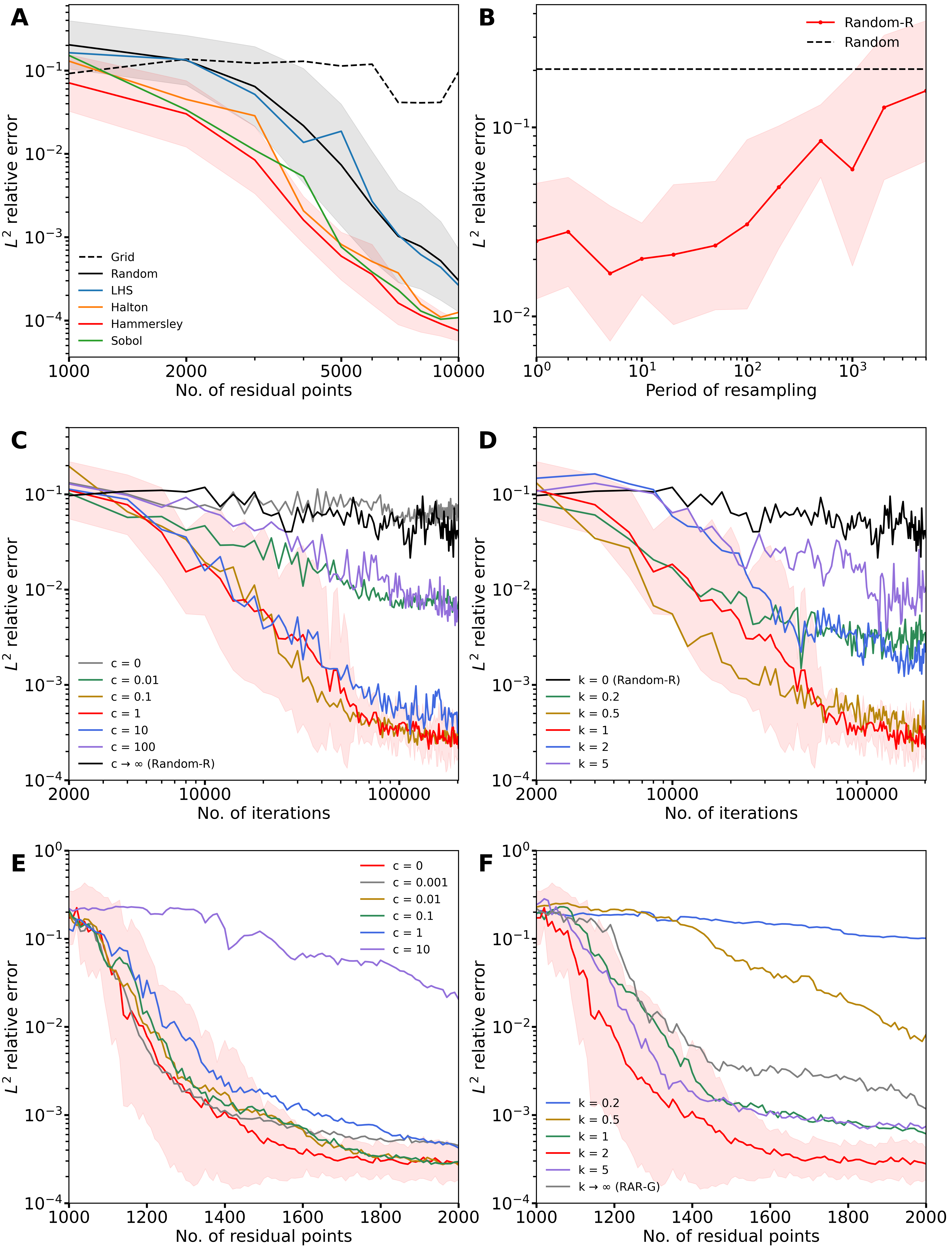}
    \caption{\textbf{$L^2$ relative errors of different sampling methods for the Burgers' equation in Section~\ref{sec:burgers}.} (\textbf{A}) Six uniform sampling with fixed residual points. (\textbf{B}) Random-R with different periods of resampling when using 2000 residual points. (\textbf{C} and \textbf{D}) The training trajectory of RAD with different values of $k$ and $c$ when using 2000 residual points. (C) $k = 1$. (D) $c = 1$. (\textbf{E} and \textbf{F}) RAR-D with different values of $k$ and $c$. Each time 10 new points are added. (E) $k$ = 2. (F) $c$ = 0. The curves and shaded regions represent the geometric mean and one standard deviation of 10 runs. For clarity, only some standard deviations are plotted.}
    \label{fig:burgers}
\end{figure}

We first test the uniform sampling methods by using the number of residual points ranging from 1,000 to 10,000 (Fig.~\ref{fig:burgers}A). The maximum iteration is 15,000 steps with Adam as optimizer followed by 15,000 steps of L-BFGS. Fig.~\ref{fig:burgers}A shows that the Hammersley method converges the fastest and reaches the lowest $L^2$ relative error among all the uniform sampling methods, while the Halton and Sobol sequences also perform adequately.

Fig.~\ref{fig:burgers}B shows the $L^2$ relative error as a function of the period of resampling using the Random-R method with 2,000 residual points. Similar to the diffusion equation, the Random-R method always outperforms the Random method. However, the performance of Random-R is not sensitive to the period of resampling if the period is smaller than 100. Choosing a period of resampling too large can negatively affect its performance.

When applying the nonuniform adaptive methods, the neural network is first trained using 15,000 steps of Adam and then 1,000 steps of L-BFGS. In the RAD method, we use 2000 residual points, which are resampled every 2,000 iterations (1,000 iterations using Adam followed by 1,000 iterations using L-BFGS). As indicated by Fig.~\ref{fig:burgers}C, the RAD method possesses significantly greater advantages over the Random-R method (a special case of RAD by choosing $k = 0$ or $c \to \infty$), whose $L^2$ relative errors barely decrease during the training processes. This fact reflects that both extreme cases show worse performance. In contrast, for $k = 1$ and $c = 1$ (the red lines in Figs.~\ref{fig:burgers}C and D), the $L^2$ relative error declines rapidly and quickly reaches $\sim 2 \times 10^{-4}$. The RAD method is also effective when choosing a set of $k$ and $c$ in a moderate range. 

For the RAR-D method, 1,000 residual points are selected in the pre-trained process, and 10 residual points are added every 2,000 iterations (1,000 iterations using Adam and 1,000 iterations using L-BFGS as optimizer) until the total number of residual points reaches 2,000. Shown by Figs.~\ref{fig:burgers}E and F, the optimal values for $k$ and $c$ are found to be 2 and 0, respectively. 

Since the solution of Burgers' equation has a very steep region, when using 2000 residual points, both RAD and RAR-D have competitive advantages over the uniform sampling methods in terms of accuracy and efficiency. For the following three forward PDE problems (Allen-Cahn equation in Section~\ref{sec:allencahn}, wave equation in Section~\ref{sec:wave}, and diffusion-reaction equation in Section~\ref{sec:diffusion-reaction}), unless otherwise stated, the maximum iterations, the use of optimizer, and the training processes remain the same as the Burgers' equation.

Table~\ref{tab:error-forward} summarizes the $L^2$ relative error for all methods when we fix the number of residual points at 2000. All uniform sampling methods fail to capture the solution well. The $L^2$ relative errors given by the Halton, Hammersley, and Sobol methods ($\sim 4\%$) are around one-fourth of that given by the Grid, Random, and LHS methods ($> 13$\%). Even though the Random-R performs the best among all uniform methods (1.69 $\pm$ 1.67\%), the proposed RAD and RAR-D methods can achieve an $L^2$ relative error two orders of magnitude lower than that (0.02\%).

\subsection{Allen-Cahn equation}
\label{sec:allencahn}

\begin{figure}[htbp]
    \centering
    \includegraphics[width=1.1\textwidth]{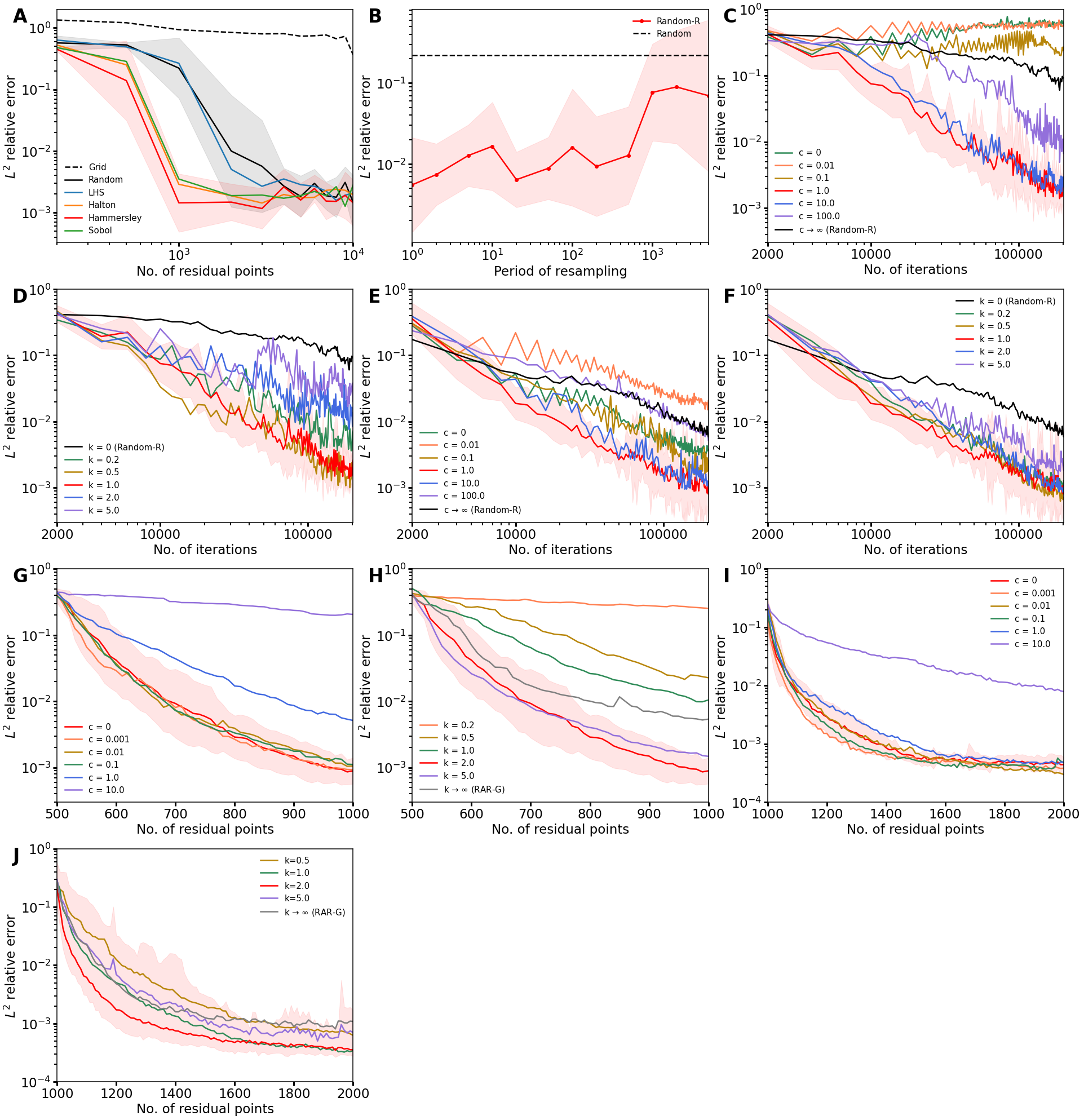}
    \caption{\textbf{$L^2$ relative errors of different sampling methods for the Allen-Cahn equation in Section~\ref{sec:allencahn}.} (\textbf{A}) Six uniform sampling with fixed residual points. (\textbf{B}) Random-R with different periods of resampling when using 1000 residual points. (\textbf{C}--\textbf{F}) The training trajectory of RAD with different values of $k$ and $c$. (C and D) 500 residual points are used. (C) $k = 1$. (D) $c = 1$. (E and F) 1000 residual points are used. (E) $k = 1$. (F) $c = 1$. (\textbf{G}--\textbf{J}) RAR-D with different values of $k$ and $c$. (G and H) The number of residual points is increased from 500 to 1000. Each time 10 new points are added. (G) $k$ = 2. (H) $c$ = 0. (I and J) The number of residual points is increased from 1000 to 2000. Each time 10 new points are added. (I) $k$ = 2. (J) $c$ = 0. The curves and shaded regions represent the geometric mean and one standard deviation of 10 runs. For clarity, only some standard deviations are plotted.}
    \label{fig:allencahn}
\end{figure}

Next, we consider the Allen-Cahn equation in the following form:
\begin{gather*} \frac{\partial u}{\partial t} = D\frac{\partial^2 u}{\partial x^2} + 5(u-u^3), \quad x \in [-1,1],t \in [0,1], \\
u(x, 0) = x^2\cos(\pi x), \\
u(-1, t) = u(1, t) = -1,
\end{gather*}
where the diffusion coefficient $D = 0.001$. Fig.~\ref{fig:allencahn} outlines the $L^2$ relative errors of different sampling methods for the Allen-Cahn equation.

Similar patterns are found for the nonadaptive uniform sampling as in the previous examples. The Hammersley method has the best accuracy (Fig.~\ref{fig:allencahn}A). As the number of residual points becomes significantly large, the difference between these uniform sampling methods becomes negligible. Except for the equispaced uniform grid method, other uniform sampling methods converge to $L^2$ relative errors of $10^{-3}$, about the same magnitude as the number of residual points reaching $10^4$. Fig.~\ref{fig:allencahn}B shows that when using 1000 residual points for Random-R, lower $L^2$ relative errors can be obtained if we select a period of resampling less than 500.

We next test the performance of RAD for different values of $k$ and $c$ when using a different number of residual points. In Figs.~\ref{fig:allencahn}C and D, we resampled 500 residual points every 2000 iteration, while in Figs.~\ref{fig:allencahn}E and F, we used 1000 residual points instead. For both cases, the combination of $k=1$ and $c=1$ (the red lines in Figs.~\ref{fig:allencahn}C--F) gives good accuracy. When fewer residual points (e.g., 500) are used, the RAD methods boost the performance of PINNs.

Similarly, we also test RAR-D in Figs.~\ref{fig:allencahn}G--J. In Figs.~\ref{fig:allencahn}G and H, we pre-train the neural network with 500 residual points and add 10 residual points after every 2000 iterations until the total number of residual points reaches 1000. In Figs.~\ref{fig:allencahn}I and J, we pre-train the neural network using 1000 residual points and heading to 2000 residual points in the same fashion. We recognize that 2 and 0 are the best $k$ and $c$ values for the RAR-D method for both scenarios, which outperform the RAR-G method.

As proven in this example, when applying the RAD and the RAR-D methods, the optimal values of $k$ and $c$ remain stable even though we choose a different number of residual points. In addition, we find that the optimal $k$ and $c$ for the Burgers' and Allen Cahn equations are the same for both the RAD and the RAR-D methods. Thus, we could choose $(k = 1, c = 1)$ for the RAD methods and  $(k = 2, c = 0)$ for the RAR-D methods by default when first applied these methods to a new PDE problem.

To make a comparison across all sampling methods, Table~\ref{tab:error-forward} shows the $L^2$ relative error for the Allen-Cahn equation when we fix the number of residual points at 1000. The Grid, Random, and LHS methods are prone to substantial errors, which are all larger than 20\%. Nevertheless, the other four uniform methods (Halton, Hammersley, Sobol, and Random-R) have greater performance and can achieve $L^2$ relative errors of less than 1\%. Remarkably, the RAD and RAR-D methods we proposed can further bring down the $L^2$ relative error below 0.1\%.

\subsection{Wave equation}
\label{sec:wave}

In this example, the following one-dimensional wave equation is considered:
\begin{gather*} 
\frac{\partial^2 u}{\partial t^2} - 4\frac{\partial^2 u}{\partial x^2} = 0, \quad x \in [0,1], t \in [0,1], \\ u(0, t) = u(1, t) = 0, \quad t\in [0,1], \\
u(x,0) = \sin(\pi x) + \frac{1}{2}\sin(4\pi x), \quad x \in [0,1], \\
\frac{\partial u}{\partial t}(x,0) = 0, \quad x \in [0,1],
\end{gather*}
where the exact solution is given as: 
\begin{equation*}
u(x,t) = \sin (\pi x) \cos (2\pi t)+ \frac{1}{2}\sin(4\pi x) \cos(8\pi t).
\end{equation*}
The solution has a multi-scale behavior in both spatial and temporal directions.

\begin{figure}[htbp]
    \centering
    \includegraphics[width=1.1\textwidth]{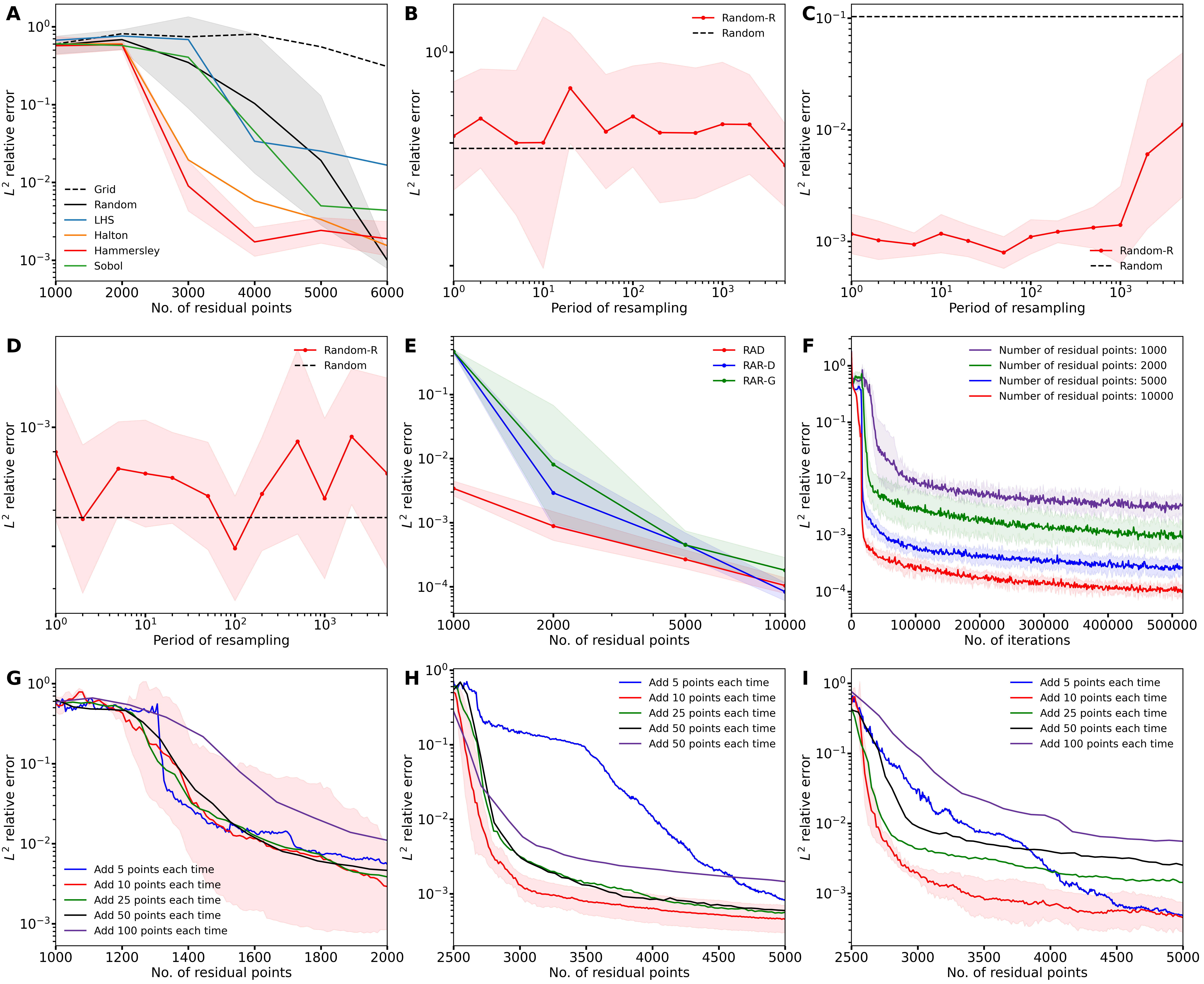}
    \caption{\textbf{$L^2$ relative errors of different sampling methods for the wave equation in Section~\ref{sec:wave}.} (\textbf{A}) Six uniform sampling with fixed residual points. (\textbf{B}, \textbf{C}, and \textbf{D}) Random-R with different periods of resampling when using (B) 1000 residual points, (C) 4000 residual points, and (D) 10000 residual points. (\textbf{E}) Comparison among RAD ($k=1$ and $c=1$), RAR-D ($k=2$ and $c=0$), and RAR-G for different numbers of residual points. (\textbf{F}) The training trajectory of RAD ($k=1$ and $c=1$) uses different numbers of residual points. (\textbf{G} and \textbf{H}) Convergence of RAR-D ($k = 2$ and $c = 0$) when adding a different number of new points each time. (G) New points are added starting from 1000 residual points. (H) New points are added starting from 2500 residual points. (\textbf{I}) Convergence of RAR-G when adding a different number of new points each time. New points are added starting from 2500 residual points. The curves and shaded regions represent the geometric mean and one standard deviation of 10 runs. For clarity, only some standard deviations are plotted.}
    \label{fig:wave}
\end{figure}

When we test the six uniform sampling methods, the number of residual points are ranged from \num{1000} to \num{6000}, with an increment of \num{1000} each time. The Hammersley method achieves the lowest $L^2$ relative error with the fastest rate (Fig.~\ref{fig:wave}A). When the number of residual points approaches \num{6000}, the Random, Halton, and Hammersley methods can all obtain an $L^2$ relative error $\sim 10^{-3}$.

To determine the effectiveness of Random-R when using different numbers of residual points, we test the following three scenarios: small (\num{1000} points), medium (\num{4000} points), and large (\num{10,000}) sets of residual points (Figs.~\ref{fig:wave}B, C, and D). In the medium case (Fig.~\ref{fig:wave}C), the Random-R attains $L^2$ relative errors magnitudes lower than the Random method. However, in the small and large cases (Figs.~\ref{fig:wave}B and D), the Random-R methods show no advantage over the Random method regardless of the period of resampling. This is because when the number of residual points is small, both the Random and Random-R methods fail to provide accurate predictions. On the other hand, if the number of residual points is large, the predictions by the Random method are already highly accurate, so the Random-R is unable to further improve the accuracy.

Since the optimal sets of $k$ and $c$ for both RAD and RAR-D methods are found to be the same for the Burgers' and the Allen Cahn equations, in this numerical experiment, we only apply the default settings (i.e., RAD: $k=1$ and $c=1$; RAR-D: $k=2$ and $c=0$) to investigate the effect of other factors, including the number of residual points for the RAD method and the number of points added to the RAR-D method.

In Fig.~\ref{fig:wave}E, we compare the performance of three nonuniform adaptive sampling methods under the same number of residual points from \num{1000} to \num{10000}. We first train the network using \num{15000} iterations of Adam and \num{1000} iterations of L-BFGS, and then after each resampling in RAD or adding new points in RAR-D/RAR-G, we train the network with \num{1000} iterations of L-BFGS. For the RAR-G and the RAR-D methods, we first train the network with 50\% of the final number of the residual points and add 10 residual points each time until reaching the total number of residual points. As we can see from Fig.~\ref{fig:wave}E, the RAD achieves much better results when the number of residual points is small. As the number of residual points increases, the RAR-D method acts more effectively and eventually reaches comparable accuracy to the RAD method. Since the RAD method is more computationally costly than the RAR-D methods with the same number of residual points, we suggest applying the RAD method when the number of residual points is small and the RAR-D method when the number of residual points is large.

We next investigate the RAD method with a different number of residual points (i.e., \num{1000}, \num{2000}, \num{5000}, and \num{10000}). Fig.~\ref{fig:wave}F illustrates that if we increase the number of residual points, lower $L^2$ relative error can be achieved but with diminishing marginal effect. We train the network for more than \num{500000} iterations to see if the $L^2$ relative error can further decrease. However, the $L^2$ relative errors converge and remain relatively stable after \num{100000} iterations.

One important factor to consider in the RAR-D and the RAR-G methods is how new points are added. We can either add a small number of residual points each time and prolong the training process or add a large number of residual points each time and shorten the process. In Fig.~\ref{fig:wave}G, we first train the network with 1000 residual points and then add new residual points at different rates until the total number of residual points reaches 2000. After adding new residual points each time, we train the network using 1000 steps of L-BFGS. Likewise, in Fig.~\ref{fig:wave}H, we first train the network with 2500 residual points and add new points at different rates until the total number of residual points reaches 5000. In both cases (Figs.~\ref{fig:wave}G and H) that use the RAR-D methods, we find that the best strategy is to add 10 points each time. However, shown by two red-shaded regions in Figs.~\ref{fig:wave}G and H, the results are more stable when we use a larger number of residual points. Fig.~\ref{fig:wave}I is set up the same way as Fig.~\ref{fig:wave}H but tests the RAR-G method. The best strategy for the RAR-G is identical to that of the RAR-D.

Table~\ref{tab:error-forward} outlines the $L^2$ relative error for the wave equation using all methods when the number of residual points equals 2000. All uniform methods with fixed residual points perform poorly (error $>50$\%) and fail to approximate the truth values. Random-R, as a special case of the proposed RAD, gives $L^2$ relative errors of around 1\%. The RAR-D method significantly enhances the prediction accuracy resulting in $L^2$ relative errors under 0.3\%. In addition, the RAD with the default setting of $k$ and $c$ converges to $L^2$ relative errors under 0.1\%.

\subsection{Diffusion-reaction equation}
\label{sec:diffusion-reaction}

The first inverse problem we consider is the diffusion-reaction system as follows:
\begin{gather*} 
\lambda\frac{d^2u}{dx^2} - k(x)u = f, \quad x\in [0, 1],
\end{gather*}
where $f = \sin(2\pi x)$ is the source term. $\lambda = 0.01$ is the diffusion coefficient, and $u$ is the solute concentration. In this problem, we aim to infer the space-dependent reaction rate $k(x)$ with given measurements on the solution $u$. The exact unknown reaction rate is
\begin{equation*}
k(x) = 0.1 + e^{-0.5\frac{(x-0.5)^2}{0.15^2}}.
\end{equation*}

\begin{figure}[htbp]
    \centering
    \includegraphics[scale=0.35]{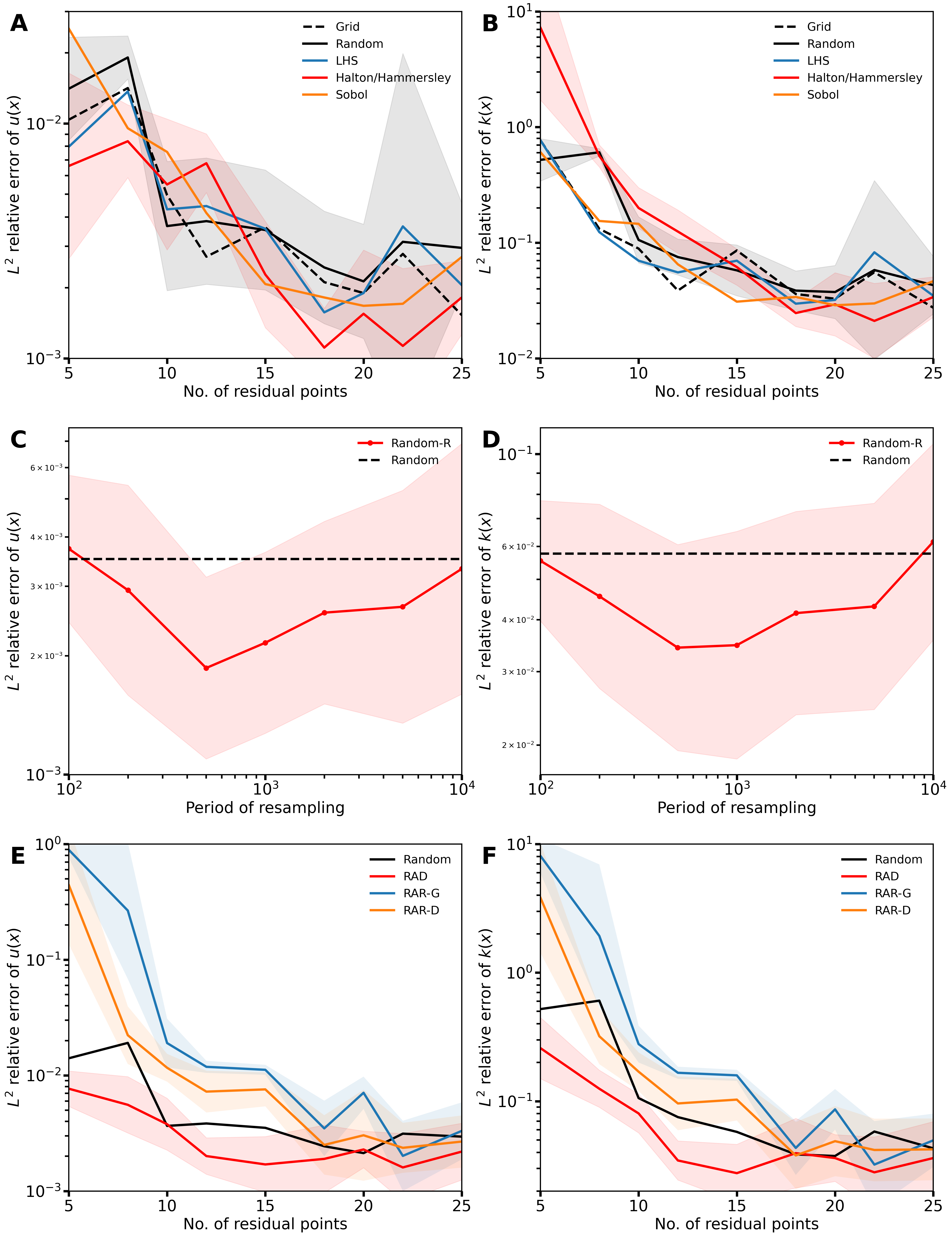}
    \caption{\textbf{$L^2$ relative errors of different sampling methods for $u$ and $k$ in the diffusion-reaction equation in Section~\ref{sec:diffusion-reaction}.} (\textbf{A} and \textbf{B}) Six uniform sampling with fixed residual points. (\textbf{C} and \textbf{D}) Random-R with different periods of resampling when using 15 residual points. (\textbf{E} and \textbf{F}) Comparison among Random, RAD ($k=1$ and $c=1$), RAR-G, and RAR-D ($k=2$ and $c=0$) for different numbers of residual points. The curves and shaded regions represent the geometric mean and one standard deviation of 10 runs. For clarity, only some standard deviations are plotted.}
    \label{fig:DR}
\end{figure}

We aim to learn the unknown function $k(x)$ and solve for $u(x)$ by using eight observations of $u$, which are uniformly distributed on the domain $x\in [0, 1]$, including two points on both sides of the boundaries. The $L^2$ relative errors for both the solution $u$ (Figs.~\ref{fig:DR}A, C, and E) and the unknown function $k$ (Figs.~\ref{fig:DR}B, D, and F) are computed. The maximum number of iterations is \num{50000} steps of Adam. Figs.~\ref{fig:DR}A and B summarize the performance of all uniform sampling methods. We note that in 1D, the Hammersley and Halton sequences are identical and outperform other uniform methods. We fix the residual points at 15 and compare the Random method with the Random-R method. The $L^2$ relative errors (Figs.~\ref{fig:DR}C and D) given by the Random-R remain steady, disregarding the changes in the period of resampling, and are approximately the same as that produced by the Random method. This is because the reaction-diffusion system is fairly simple and can be easily handled by uniform sampling methods without resampling.

Next, we compare the Random, RAD, RAR-G, and RAR-D methods with default settings (i.e., RAD: $k=1$ and $c=1$; RAR-D: $k=2$ and $c=0$) using a different number of residual points. For the random and RAD methods, the maximum number of iterations is \num{50000} steps of Adam. For the RAR-G/RAR-D, we first train the neural network with 50\% of the total number of residual points for \num{10000} steps of Adam; then we add one point each time and train for \num{1000} steps of Adam until we meet the total number of residual points. As shown by Figs.~\ref{fig:DR}E and F, the RAD method surpasses other methods and is able to produce low $L^2$ relative error even when the number of residual points is very small. However, RAR-G and RAR-D are even worse than the Random sampling.

To sum up, we fix the number of residual points at 15 and present the $L^2$ relative error for both the solution and unknown function in Table~\ref{tab:error-inverse}. The RAD yields the minimum $L^2$ relative error (0.17\% for $u(x)$; 2.76\% for $k(x)$). However, due to the simplicity of this PDE problem, some uniform sampling methods, especially the Sobol and Random-R, have comparable performance to the RAD. Generally speaking, we recognize that the uniform sampling methods are adequate when solving this inverse PDE with smooth solutions. Still, the RAD method can further enhance the performance of PINNs, especially when the number of residual points is small.

\subsection{Korteweg-de Vries equation}
\label{sec:kdv}

The second inverse problem we solve is the Korteweg-de Vries (KdV) equation:
\begin{gather*}
\frac{\partial u}{\partial t} + \lambda_{1}u\frac{\partial u}{\partial x} + \lambda_{2}\frac{\partial^{3} u}{\partial x^3} = 0,  \qquad x\in [-1, 1], \quad t\in [0, 1],
\end{gather*}
where $\lambda_{1}$ and $\lambda_{2}$ are two unknown parameters. The exact values for $\lambda_1$ and $\lambda_2$ are 1 and 0.0025, respectively. The initial condition is $u(x,t=0) = \cos (\pi x)$, and periodic boundary conditions are used. To infer $\lambda_1$ and $\lambda_2$, we assume that we have the observations of two solution snapshots $u(x,t=0.2)$ and $u(x,t=0.8)$ at 64 uniformly distributed points at each time.

\begin{figure}[htbp]
    \centering
    \includegraphics[width=1.1\textwidth]{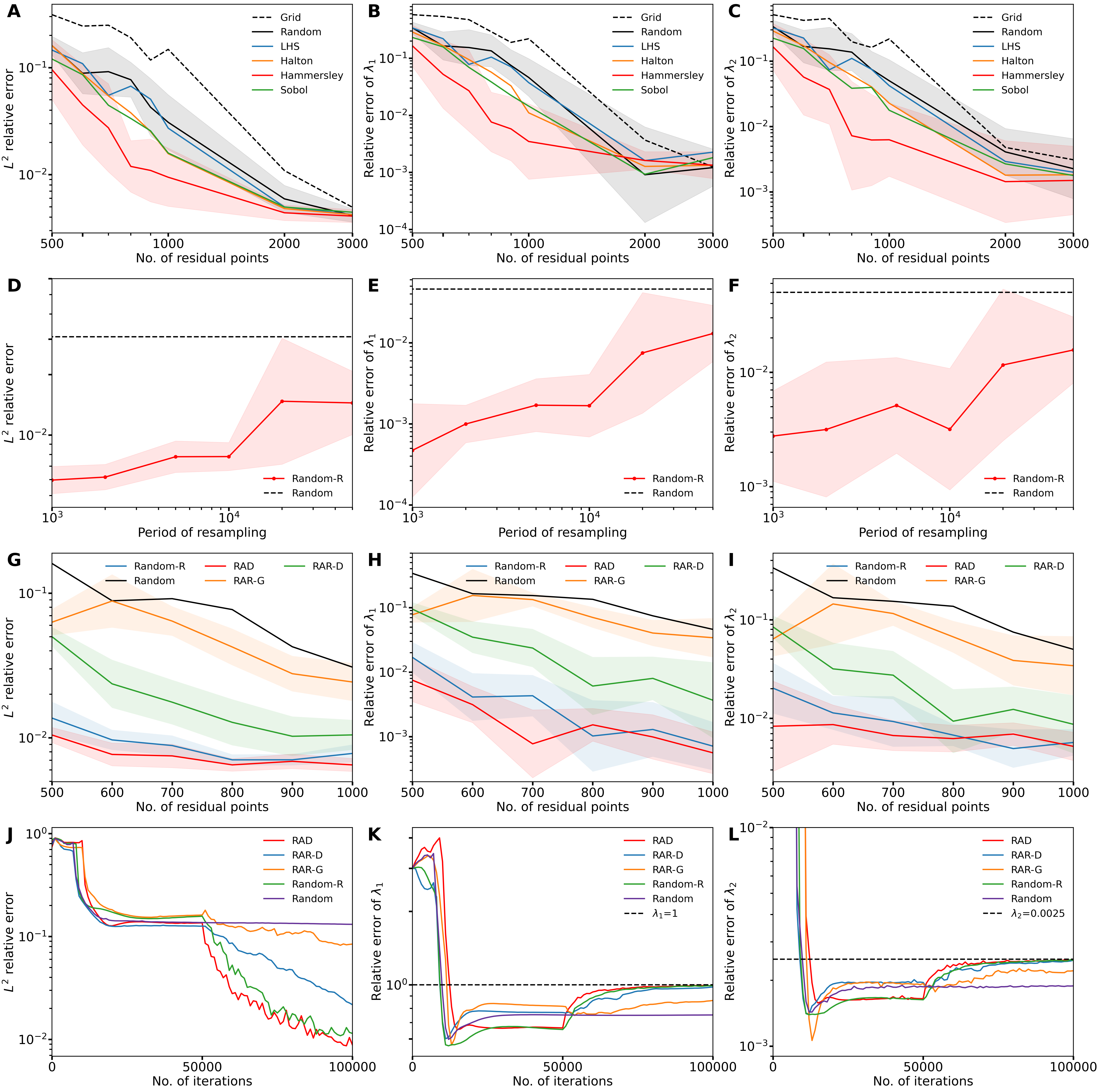}
    \caption{\textbf{$L^2$ relative errors of $u$ and relative errors of $\lambda_1$ and $\lambda_2$ using different sampling methods for the Korteweg-de Vries equation in Section~\ref{sec:kdv}.} (\textbf{A}, \textbf{B}, and \textbf{C}) Six uniform sampling with fixed residual points. (\textbf{D}, \textbf{E}, and \textbf{F}) Random-R with different periods of resampling when using 1000 residual points. (\textbf{G}, \textbf{H}, and \textbf{I}) Comparison among Random, Random-R, RAD ($k=1$ and $c=1$), RAR-G, and RAR-D ($k=2$ and $c=0$) for different number of residual points. (\textbf{J}, \textbf{K}, and \textbf{L}) Examples of the training trajectories using Random, Random-R, RAD ($k=1$ and $c=1$), RAR-G, and RAR-D ($k=2$ and $c=0$) with 600 residual points. The curves and shaded regions represent the geometric mean and one standard deviation of 10 runs. For clarity, only some standard deviations are plotted.}
    \label{fig:kdv}
\end{figure}

In Fig.~\ref{fig:kdv}, the first column (Figs.~\ref{fig:kdv}A, D, and G) shows the $L^2$ relative error of the solution $u$, while the second column (Figs.~\ref{fig:kdv}B, E, and H) and the third column (Figs.~\ref{fig:kdv}C, F, and I) illustrate the relative errors for $\lambda_1$ and $\lambda_2$, respectively. The maximum iteration is \num{100000} steps of Adam. Hammersley achieves better accuracy than the other uniform sampling methods. The Sobol and Halton methods behave comparably as these two curves (the yellow and green curves in Figs.~\ref{fig:kdv}A, B, and C) are almost overlapping. Shown in Figs.~\ref{fig:kdv}D, E and F, the Random-R method yields higher accuracy than the Random method by about one order of magnitude in all cases when using 1000 residual points. A smaller period of resampling leads to smaller errors.

Figs.~\ref{fig:kdv}G, H, and I compare the Random-R, Random, RAD, RAR-G, and RAR-D methods using the same number of residual points and the total number of iterations. For the Random and the Random-R methods, we train the network for \num{100000} steps of Adams. For the RAD methods, we first train the network using \num{50000} steps of Adams; then, we resample the residual points and train for 1000 steps of Adams 50 times. In order to fix the total number of iterations for the RAR-G/RAR-D methods to \num{100000}, we accordingly adjust the number of new residual points added each time. For example, if the final number of residual points is 500, we first train the network using 250 residual points (i.e., 50\% of the total number of residual points) with \num{50000} steps of Adams; and we consequently add 5 points and train for 1000 steps of Adams each time. If the final number of residual points is 1000, we first train the network using 500 residual points with \num{50000} steps of Adams; and then we add 10 points and train for 1000 steps of Adams each time. As demonstrated by Figs.~\ref{fig:kdv}G, H, and I, the RAD method is the best, while the Random-R method is also reasonably accurate. We show one example of the training process (Figs.~\ref{fig:kdv}J, K, and L) when the number of residual points is 600 to illustrate the convergence of the solution, $\lambda_1$, and $\lambda_2$ during training. The resampling strategies, especially the RAD method, achieve the greatest success among all sampling methods.

Table~\ref{tab:error-inverse} demonstrates the $L^2$ relative errors for the solution $u(x,t)$ and the relative error of two unknown parameters $\lambda_1$ and $\lambda_2$, for all methods when the number of residual points is set at 600. The lowest $L^2$ relative errors for uniform sampling with fixed points are given by Hammersley ($\sim 5\%$). The Random-R is the second-best method and provides $L^2$ relative errors of around 1\%. With the smallest errors ($<1\%$) and standard deviations, the RAD method has compelling advantages over all other methods in terms of accuracy and robustness. It is noteworthy that the RAR-D method provides adequate accuracy ($\sim 3\% $) and is less expensive than the Random-R and RAD methods when the number of residual points is the same. Therefore, the RAR-D is also a valuable approach to consider.

\section{Conclusions}
\label{sec:conclusion}

In this paper, we present a comprehensive study of two categories of sampling for physics-informed neural networks (PINNs), including non-adaptive uniform sampling and adaptive nonuniform sampling. For the non-adaptive uniform sampling, we have considered six methods: (1) equispaced uniform grid (Grid), (2) uniformly random sampling (Random), (3) Latin hypercube sampling (LHS), (4) Halton sequence (Halton), (5) Hammersley sequence (Hammersley), and (6) Sobol sequence (Sobol). We have also considered a resampling strategy for uniform sampling (Random-R). For the adaptive nonuniform sampling, motivated by the residual-based adaptive refinement with greed (RAR-G)~\cite{lu2021deepxde}, we proposed two new residual-based adaptive sampling methods: residual-based adaptive distribution (RAD) and residual-based
adaptive refinement with distribution (RAR-D).

We extensively investigated the performance of these ten sampling methods in solving four forward and two inverse problems of partial differential equations (PDEs) with many setups, such as a different number of residual points. Our results show that the proposed RAD and RAR-D significantly improve the accuracy of PINNs by orders of magnitude, especially when the number of residual points is small. RAD and RAR-D also have great advantages for the PDEs with complicated solutions, e.g., the solution of the Burgers' equation with steep gradients and the solution of the wave equation with a multi-scale behavior. A summary of the comparison of these methods can be found in Section~\ref{sec:summary}.

Based on our empirical results, we summarize the following suggestions as a practical guideline in choosing sampling methods for PINNs.
\begin{itemize}
    \item RAD with $k=1$ and $c=1$ can be chosen as the default sampling method when solving a new PDE. The hyperparameters $k$ and $c$ can be tuned to balance the points in the locations with large and small PDE residuals.
    \item RAR-D can achieve comparable accuracy to RAD, but RAR-D is more computationally efficient as it gradually increases the number of residual points. Hence, RAR-D ($k=2$ and $c=0$ by default) is preferable for the case with limited computational resources.
    \item Random-R can be used in the situation where adaptive sampling is not allowed, e.g., it is difficult to sample residual points according to a probability density function. The period of resampling should not be chosen as too small or too large.
    \item A low-discrepancy sequence (e.g., Hammersley) should be considered rather than Grid, Random, or LHS, when we have to use a fixed set of residual points, such as in PINNs with the augmented Lagrangian method (hPINNs)~\cite{lu2021physics}.
\end{itemize}

In this study, we sample residual points in RAD and RAR-D by using a brute-force approach, which is simple, easy to implement, and sufficient for many PDEs. However, for high-dimensional problems, we need to use other methods, such as generative adversarial networks (GANs)~\cite{goodfellow2014generative}, as was done in Ref.~\cite{tang2021deep}. Moreover, the probability of sampling a point $\mathbf{x}$ is only considered as $p(\mathbf{x}) \propto \frac{\varepsilon^k(\mathbf{x})}{\mathbb{E}[ \varepsilon^k(\mathbf{x})]} + c$. While this probability works very well in this study, it is possible that there exists another better choice. We can learn a new probability density function by meta-learning, as was done for loss functions of PINNs in Ref.~\cite{psaros2022meta}.

\bibliographystyle{unsrt}
\bibliography{main}

\end{document}